\documentclass[a4paper,twocolumn]{esaproc} 
\usepackage{authordate1-4}
\usepackage{epsf}
\title{Interstellar Dust Module for the ESA Meteoroid Model}
\author[1]{M. Landgraf}
\author[1]{R. Jehn}
\author[1]{N. Altobelli}
\affil[1]{ESA/ESOC, Robert-Bosch-Str. 5, 64293 Darmstadt, Germany}
\author[2]{V. Dikarev}
\author[2]{E. Gr\"un}
\affil[2]{MPI-K, Postfach 103980, 69029 Heidelberg, Germany}
\begin{document}
\maketitle
\begin{abstract}
The ESA meteoroid model predicts impacts of meteoroids in the mass
range between $10^{-18}$ to $10^0\:{\rm g}$ on spacecraft surfaces. It
covers heliocentric distances from $0.3$ to $20\:{\rm
AU}$. Measurements of the dust detector on board the highly successful
joint ESA/NASA mission Ulysses have shown, that the flux of meteoroids
with masses between $10^{-15}$ and $10^{-12}\:{\rm g}$ is, at least in
the outer Solar System, dominated by interstellar dust grains that
traverse the Solar System as it travels through the local interstellar
cloud. We present a simple semi-analytic interstellar dust model that
can easily be included in the ESA meteoroid model, together with a
more precise determination of the flux direction of the interstellar
dust stream. The model is based on the assumption that interstellar
dust dynamics have two effects: solar gravitation and radiation
pressure determines the spatial distribution, and Lorentz-interaction
of the charged particles creates a temporal variation.
\end{abstract}

\section{Introduction}
As meteoroid impacts are a concern for highly sensitive instruments on
board Earth orbiting satellites as well as interplanetary exploration
probes, ESA has decided to develop a standard tool for the prediction
of impact fluxes. In its first release version the tool
\cite{staubach97} has been used to predict meteoroid impact rates in
order to assess its contribution relative to the small debris
population in Earth orbit. Due to the increase of data available from
the in situ measurements of Ulysses and Galileo, as well as from the
radar meteor measurements by the AMOR facility in New Zealand, ESA has
decided to upgrade the tool. In the course of this update, an
extension of the interstellar dust module was proposed.

\begin{figure}[ht]
\center
\epsfbox{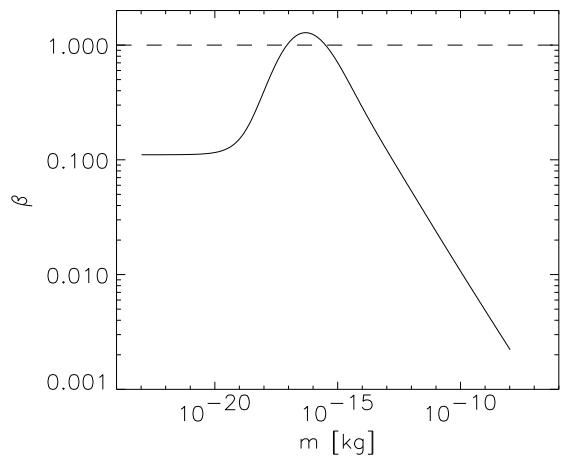}
\caption{\label{fig_bobeta} Ratio $\beta$ of radiation pressure force
to gravity on dust grains in the Solar System as a function of grain
mass $m$. The dashed line indicates $\beta=1$, for which radiation
pressure and gravity exactly cancel each other
\protect\cite{gustafson94}.}
\end{figure}

The former implementation of the ESA meteoroid model includes an
interstellar dust component that is modelled as a constant,
mono-directional stream of particles moving towards a heliocentric
ecliptic direction of $\beta_{\rm ECL}=-5^o$, $\lambda_{\rm
ECL}=79^o$, with a flux density (integrated over all grain sizes) of
$1.5\times 10^{-4}\:{\rm m}^{-2}\:{\rm s}^{-1}$ and a velocity of
$26\:{\rm km}\:{\rm s}^{-1}$.The direction is identical to the
downstream direction of interstellar neutral Helium on its way through
the heliosphere
\cite{witte93}. From the analysis of the Ulysses dust data it was
found, that the downstream direction of interstellar dust is
statistically indistinguishable from the gas direction
\cite{frisch99}. The current implementation of the interstellar dust
module assumes that the particles move on straight lines through the
Solar System. This is a good approximation for particles with masses
around $3\times 10^{-16}\:{\rm kg}$, for which the forces of gravity
and radiation pressure nearly cancel each other. The ratio $\beta$ of
radiation pressure to gravity is equal unity for these particles (see
figure \ref{fig_bobeta}).

\section{Spatial Distribution of Interstellar Dust in the Solar System}

\begin{figure}[ht]
\center
\epsfxsize=.8\hsize
\epsfbox{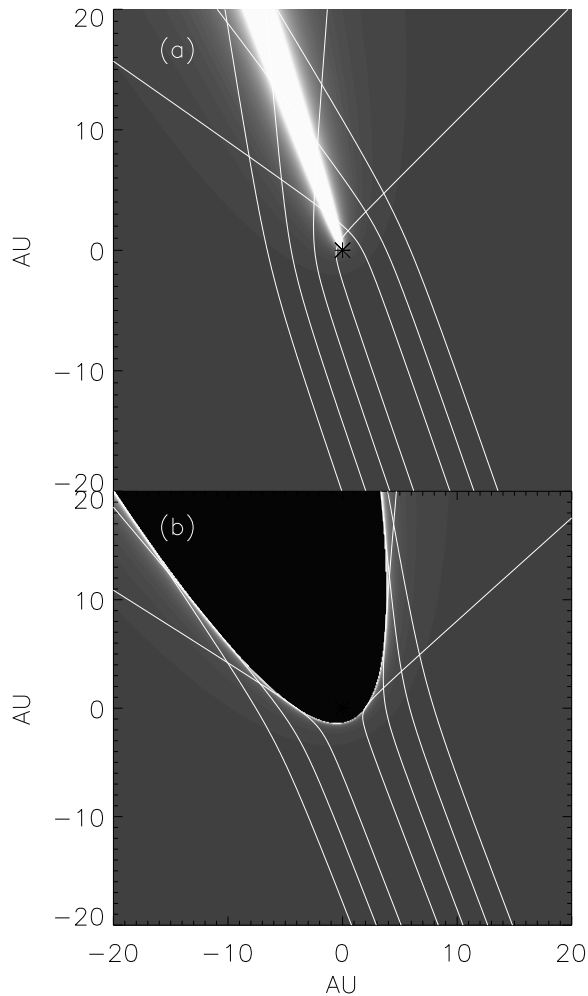}
\caption{\label{fig_distributions} Spatial distribution and motion of
interstellar dust grains with $\beta \neq 1$ in a $40\times 40\:{\rm
AU}$ region around the Sun in the ecliptic plane. The gray-scale
contours indicate the spatial concentration and the solid lines show
the hyperbolic paths of grains with selected initial
conditions. Diagram (a) shows the situation for grains with
$\beta=0.5$, and diagram (b) for grains with $\beta=1.5$.}
\end{figure}

For values of $\beta$ other than unity, the particles move on Kepler
hyperbola (figure \ref{fig_distributions}). As a consequence, the
apparent stream direction and local concentration depends on the
location of the detector in the solar system \cite{landgraf99c}.

\section{Temporal Modulation of the Local Interstellar Flux}

\begin{figure}[ht]
\center
\epsfbox{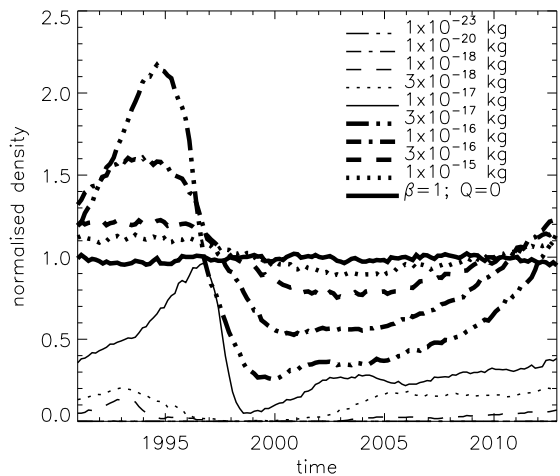}
\caption{\label{fig_tempvariation} Temporal variation of the flux of
interstellar dust in the solar system (inside $10\:{\rm AU}$) over a
full $22\:{\rm year}$ solar cycle for various grain masses. As a
reference, grains with $\beta=1$ and charge $Q=0\:{\rm C}$ are given
(thick, solid line) that are effectively not affected by gravity,
radiation pressure, and Lorentz-force.}
\end{figure}

Ulysses measured a variation of a factor of $3$ in the flux of
interstellar dust grains in the Solar System \cite{landgraf00a}. This
was attributed to the electromagnetic interaction of the charged
grains with the solar wind magnetic field that varies with the
$22\:{\rm year}$ solar cycle \cite{landgraf00}. Interstellar dust
grains are charged mainly due to the photo-effect caused by solar UV
photons. Because of the magnetic field that is frozen in the radially
expanding solar wind (speed between $300$ and $700\:{\rm km}\:{\rm
s}^{-1}$), a Lorentz force acts on the grains. The direction this
force depends on the phase of the solar cycle. During the period from
1991 to 2002 it is was directed such that interstellar grains are
deflected away from the solar equator. Consequently the local
interstellar dust concentration is significantly reduced. While proof
of the electromagnetic hypothesis still pending, we employ the model
described by Landgraf \shortcite{landgraf00} to calculate the temporal
flux variation for various grain sizes (and thus charge-to-mass
ratios). Figure \ref{fig_tempvariation} shows the normalised spatial
concentration of grains of various sizes as a function of time over
one full solar cycle from 1991 to 2013. For flux predictions after
2013 we assume a perfect periodicity of the solar cycle.

Because of the differential susceptibility of small and big grains to the
electromagnetic deflection described above, the grain mass
distribution far from the Sun can not me measured directly inside the
solar system. We assume that the grain masses are distributed
according to
\begin{eqnarray}
n(m)dm & = & n_0 m^{-/1.83}dm.
\end{eqnarray}
We implement the temporal variation of the dust concentration in the
interstellar dust module of the ESA meteoroid model by modulating the
grain mass distribution over time.

\section{Interstellar Dust Stream Direction}
The interstellar dust stream direction has been determined from
Ulysses data collected within four weeks of the Jupiter fly-by
\cite{frisch99}. With the now much extended database, a more accurate
determination is possible.

\begin{figure}[ht]
\center
\epsfxsize=\hsize
\epsfbox{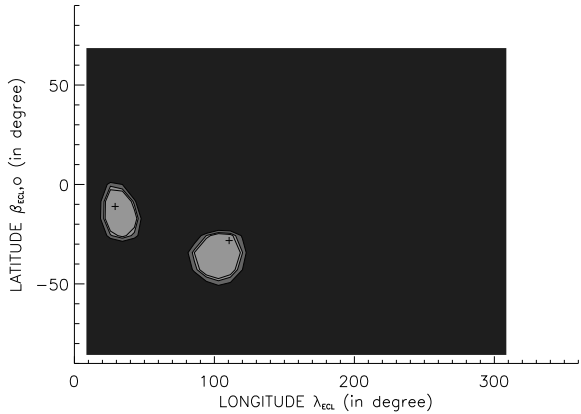}
\caption{\label{fig_chisquared} Best-fit $\chi^2$ analysis of the
Ulysses data between Jupiter fly-by and south polar pass. The contour
levels (from light to dark) indicate $10\%$, $5\%$, and $1\%$
confidence intervals. Crosses mark local minima of the $\chi^2$
function. The lower right cross represents the global minimum ($99\%$
confidence level), the second local minimum represents a fit with a
$98\%$ confidence level.}
\end{figure}
Figure \ref{fig_chisquared} shows the fit-levels of an interstellar
dust directional model compared with data provided by the Ulysses dust
experiment in the time range from the Jupiter fly-by in February 1992
to January 1994, just before the first pass of Ulysses over the south
pole of the Sun. For the analysis we use only impacts with a mass
(also measured by the detector) between $10^{-14}\:{\rm kg}$ and
$10^{-12}\:{\rm kg}$. For these grains the solar radiation pressure
and the gravity of the Sun are equal (figure \ref{fig_bobeta}). Thus,
it can be assumed that such grains are not deviated from their
original direction of motion. Moreover, only impacts, which are the
most reliably detected, are used for the comparison with the
model. The interstellar dust flow model can be characterised by the
vector $(\lambda_{\rm ECL},\beta_{{\rm ECL},0},\sigma_\beta)$, where
$\lambda_{\rm ECL}$ and $\beta_{{\rm ECL},0}$ are respectively the
ecliptic longitude and latitude of the dust flow direction, and
$\sigma_\beta$ the width of the directional distribution. From this
model the total number of dust impacts for each spacecraft rotation
angle interval is calculated. The expected number of impacts per
rotation angle, derived from the model and the measured number of
impacts per rotation angle interval have been compared by the $\chi^2$
statistical method. Its minimum, presented in figure
\ref{fig_chisquared}, gives the model parameters providing the best
fit to the data. The best-fit parameters for the main downstream
direction of interstellar dust in the Solar System is heliocentric
longitude $\lambda_{\rm ECL}=110^\circ$ and heliocentric latitude
$\beta_{{\rm ECL},0}=-30^\circ$. Also a broadening of the
directionality by $\sigma_\beta=20^\circ$ was found to provide the
best fit.

It can be seen from figure \ref{fig_chisquared} that the best-fit
flow direction of interstellar dust is not determined
unambiguously. Another good fit is achieved for $\lambda_{\rm ECL} =
30^\circ$ and $\beta_{{\rm ECL},0} = -10^\circ$. Analysis of the
Ulysses data is ongoing in order to resolve this ambiguity.

\bibliography{debris,mas,dust}
\bibliographystyle{authordate1}

\end{document}